\documentclass[a4paper,10pt,twoside]{cpc-hepnp}
\usepackage{multicol}
\usepackage{graphicx}
\usepackage{booktabs}
\usepackage{amssymb,bm,mathrsfs,bbm,amscd}
\usepackage[tbtags]{amsmath}
\usepackage{lastpage}
\usepackage{CJK}\usepackage{epsfig}

\begin{document}
\begin{CJK*}{GBK}{song}

\title{Three-body force effect on the properties of nuclear matter under the gap and continuous choices within the BHF approach}

\author{ Wang Pei$^{1,2,3}$ and Zuo Wei$^{1,4,5;1)}$\email{zuowei@impcas.ac.cn} }

\maketitle

\address{1~(Institute of Modern Physics, Chinese Academy of Sciences, Lanzhou
730000, China)\\
2~(School of Physical Science and Technology, Lanzhou University, Lanzhou 730000, China)\\
3~(Graduate School of Chinese Academy of Sciences, Beijing 100039, China)\\
4~(Kavli Institute for Theoretical Physics China, Chinese Academy of Sciences, Beijing 100190, China)\\
5~(State Key Laboratory of Theoretical Physics,
 Institute of Theoretical Physics, Chinese Academy of Sciences, Beijing 100190, China)}
\begin{abstract}
We have calculated and compared the three-body force effects on the properties of nuclear matter
under the gap and continuous choices for the self-consistent auxiliary potential
within the Brueckner-Hartree-Fock approach
by adopting the Argonne $V_{18}$ and the Bonn B
two-body potentials plus a microscopic three-body force (TBF). The
TBF provides a strong repulsive effect on the equation of state of nuclear matter at high densities
for both the gap and continuous choices. The saturation point turns out to be much closer to the empirical
value when the continuous choice is adopted. In addition, the
dependence of the calculated symmetry energy upon the choice of
the self-consistent auxiliary potential is discussed.
\end{abstract}



\begin{multicols}{2}

\section{Introduction}
One of the original aims of the microscopic
Bethe-Brueckner-Goldstone (BBG) theory of nuclear matter
\cite{ba1} is to study the equation of state (EoS) of nuclear matter and reconcile the empirical saturation
point. There are two most important uncertain points in the BBG theory
\cite{ba1}. One is the choice of the auxiliary potential $U(k)$,
another is relevance of the higher order contributions in the BBG
expansion. Over the last decades, the many-body uncertainties in
the BBG theory have been checked carefully, and considerable progress has been made for improving the predicted
saturation point.

In an early stage of the BBG theory \cite{bru1,koh1,day3}, the
auxiliary potential $U(k)$ is introduced based on the re-summation
of Feynman diagrams \cite{koh1}, in order to decrease the number of
diagrams in calculation \cite{bru1,day3}. It has been argued that
different choices for the self-consistent auxiliary potential have
effect on the convergence rate of the hole-line expansion, and can
produce energy shifts in the calculated total binding energy \cite{jp1}. As
shown by Day and Wiringa \cite{day1,day2}, the three-hole line
contribution in the BBG expansion is non-negligible in the gap choice
of the auxiliary potential. Baldo and Song have extended the analysis to
symmetry nuclear matter \cite{ba2,ba3,ba4} and pure neutron
matter \cite{ba5} at the three-hole line level with the local
separable Argonne $V_{14}$ (\emph{AV}14) \cite{av1} and Argonne $V_{18}$ (\emph{AV}18) \cite{av2}
two-nucleon potentials by solving the Bethe-Fadeev equations \cite{bru1,be1}.
Their results give a strong evidence that
convergence has been reached \cite{ba6,ba7} within the gap and
continuous choices for density up to six times the saturation
value \cite{ba8}. Furthermore, the results show that using the
continuous choice leads to a faster convergence than the gap choice \cite{ba2}.
Since the EoS of nuclear matter are obtained at the level of two-hole line,
the results of Ref.\cite{ba2} indicates that the Brueckner-Hartree-Fock (BHF) approximation under the
continuous choice is able to incorporate large part of the
three-hole correlations within the gap choice.
Although the convergence can be reached,
the BHF approach is not able to reproduce the empirical saturation point by adopting purely two-body interactions in both the
gap and the continuous choices \cite{ba2,ba8,ba3}.
Three-body forces (TBF) are required for describing reasonably the nuclear saturation properties \cite{tbf1,tbf2,tbf3,lizh:2008}.

In recent years, calculations in the BHF approach have been updated
by incorporating the consistent microscopic TBFs
\cite{tbf1,tbf2,tbf3,lizh:2008} constructed consistently with the realistic Bonn B \cite{bonn} and
\emph{AV}18 two-nucleon potentials as input. As shown in
Refs. \cite{tbf3,lizh:2008}, the saturation points can be improved by including the TBFs from
($0.34 $ fm$^{-3}$, $-22$ MeV) to ($0.17 $ fm$^{-3}$, $-15.9$ MeV) and from
($0.27 $ fm$^{-3}$, $-18.3$ MeV) to ($0.2 $ fm$^{-3}$, $-15.1$ MeV), for the Bonn B and
\emph{AV}18 potentials respectively, under the continuous
choice.

In this work, we extend the previous results within the BHF
approach in the gap and continuous choices by employing the
Bonn B and \emph{AV}18 potentials plus their corresponding self-consistent TBFs,
respectively. We concentrate on the comparison between the TBF effects on the properties of nuclear matter obtained under the gap choice
and the continuous choice. The symmetry
energy within the two different choices for the self-consistent auxiliary
potential is also investigated and discussed. In Sec. II, we give a brief review of the BHF
theory and the TBF model. The numerical results are presented and discussed in Sec. III.
Our conclusions are summarized in the last section.

\section{Formalism}
Our calculations are based on the microscopic
BHF approach. The BHF description of nuclear
matter is derived by a linked cluster of independent hole-line
expansion. The starting point of this theory is the in-medium two-body
Brueckner reaction matrix $G$, which is the solution of the
Bethe-Goldstone equation:
\begin{eqnarray}
G(\rho;\omega)&=&v+
v\sum\limits_{k_{1}k_{2}}\frac{|k_{1}k_{2}\rangle
Q(k_{1},k_{2})\langle
k_{1}k_{2}|}{\omega-\epsilon(k_{1})-\epsilon(k_{2})+i\eta}G(\rho;\omega)
\nonumber\\
\end{eqnarray}
where $v$ is the realistic nucleon-nucleon (NN) interaction,
$\omega$ is the starting energy, and $\rho$ denotes nucleon number density.
$Q(k_{1},k_{2})$ is the Pauli operator, which prevents the two
nucleons from being scattered into their respective Fermi-seas.
The single-particle (s.p.) energy is given by:
$\epsilon(k)=\frac{\hbar^{2}k^{2}}{2m}+U(k)$, where $U(k)$ is the auxiliary s.p. potential.
Within the framework of the BHF approximation, the convergence rate of the hole-line
expansion depends on the specified choice of the auxiliary potential \cite{ba1}.
Two different choices have been usually adopted in the BHF calculations \cite{ba1}:
one is the continuous choice, the other is the gap choice. Under the continuous choice,
the auxiliary potential is given by:
\begin{eqnarray}
U(k)={\rm Re} \sum\limits_{k'\leq k_{F}}\langle
kk'|G[\rho,\epsilon(k)+\epsilon(k')]|kk'\rangle_{A}
\label{eq:u}\end{eqnarray}
where the subscript $A$ denotes antisymmetrization of the matrix
elements. For the gap choice, the auxiliary potential for the hole states $(k<k_{F})$ is calculated according to Eq.(\ref{eq:u}),
 while it is set to zero above the Fermi surface $(k>k_{F})$.
There are also other possibilities for choosing the auxiliary potential, e.g. the model-space BHF (MBHF) \cite{mod}.
In this work, we will
restrict the calculations to the gap and continuous choices.

For the realistic NN interaction, we adopt different realistic two-body interactions (i.e., the $AV18$ and the Bonn B potentials~\cite{bonn}) plus their corresponding microscopic TBFs which are base on the meson exchange current approach \cite{tbf1,tbf2,tbf3,lizh:2008}.
In the OBEP approximation, $\pi$, $\rho$,
$\sigma$, $\omega$ mesons are considered and the corresponding meson
parameters (meson-nucleon couplings and form factors) in the TBF model are
determined self-consistently with the two-body potentials. In
the present calculation, the TBF is reduced to an equivalently
effective two-body force by averaging over the third nucleon
degree of freedom in nuclear medium \cite{tbf1,tbf4,tbf5}:
\begin{eqnarray}
\overline{V}_{ij}(r)=\rho\int
d^{3}r_{k}\sum\limits_{\sigma_{k},\tau_{k}}[1-g(r_{ik})]^{2}[1-g(r_{jk})]^{2}V_{ijk}
\end{eqnarray}
where $g(r)$ is the defect function which reflects the
NN correlations in nuclear medium.
 Within the BHF approximation, the energy per nucleon is given by:
\begin{eqnarray}
\frac{B}{A}=\frac{3}{5}\frac{k_{F}^{2}}{2m}+\frac{1}{2\rho}{\rm Re}\sum_{k,k'\leq
k_{F}}\langle kk'|G[\rho;\epsilon(k)+\epsilon(k')]|kk'\rangle_{A}
\nonumber
\end{eqnarray}

\section{Results and discussion}
Fig. \ref{fig1} shows the self-consistent s.p. potentials in symmetric
nuclear matter at density $\rho=0.17$fm$^{-3}$, obtained respectively under
the gap choice (BHFG) and the continuous choice (BHFC). For the continuous
choice, it is seen that the s.p. potential is strongly attractive
at low momenta and its attraction decreases monotonically as a
function of momentum continuously through the Fermi surface. In the
gap choice, the self-consistent s.p.
potential is cut off for momenta $k$ larger than $k_{F}$, and
shows a big gap at the Fermi surface. It is clearly from Fig. \ref{fig1}
that, the s.p. potentials are more attractive in the continuous
choice than those in the gap choice, indicating that the effective
interaction is more strongly attractive between the nucleons in
the continuous choice than the gap choice. This is consistent
with the analysis in Refs.\cite{jp1,ba2,be2} where it has been shown that the
suppression of the gap at $k_{F}$ in the s.p. potential tends to
increase the correlation energy.
\begin{center}
\includegraphics[width=8cm]{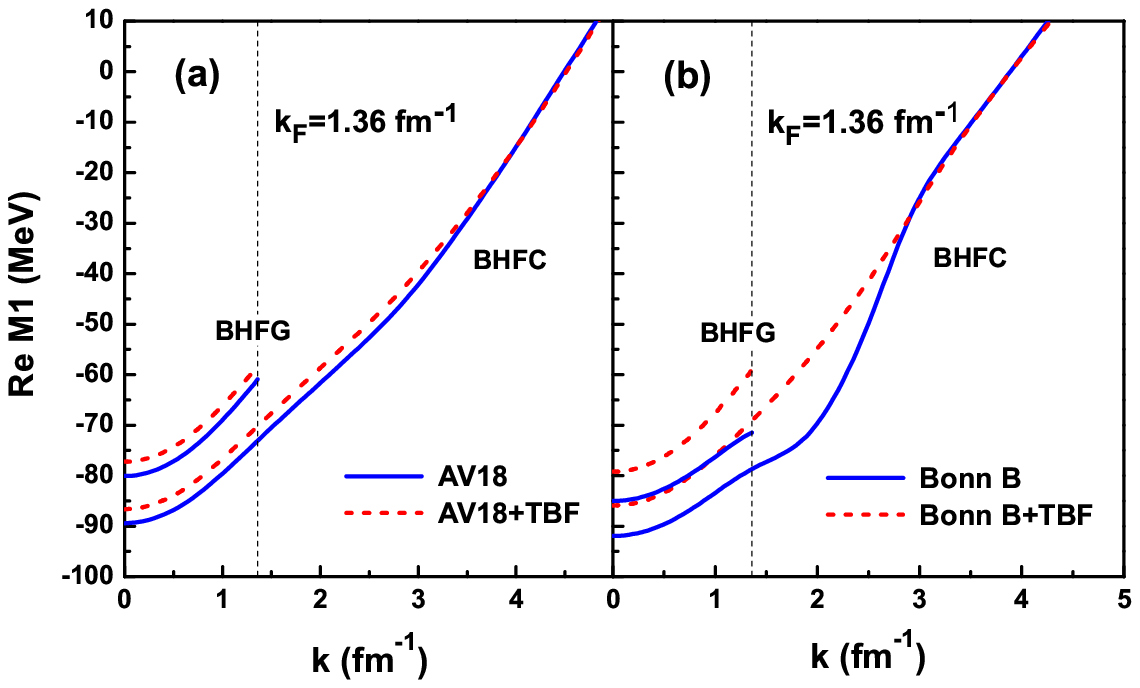}
\figcaption{(color online) (a) The s.p.
potential in symmetric nuclear matter at $\rho=0.17$fm$^{-3}$, obtained with the \emph{AV}18 potential. The dashed
(red) curves have been obtained by taking into account the TBF contribution, while
the solid (blue) curves denote the results without including the TBF. The vertical dashed lines show the
location of the Fermi surface. The s.p. potentials
discontinuous and continuous at the Fermi momentum $k_{F}$
correspond to the gap (BHFG) and continuous choice (BHFC),
respectively. (b) The same as in (a), but for the Bonn B
potential.}\label{fig1}
\end{center}

The TBF effects on the s.p. potential are also reported in Fig. \ref{fig1}
(dashed lines) with the gap choice and the continuous choice,
respectively. We find that the TBF leads to a repulsive
contribution and affects the s. p. potential mainly at low momentum
region not only for the Bonn B potential (right panel) but also
for \emph{AV}18 potential (left panel), besides the less
magnitude for the \emph{AV}18 potential, both in the gap and the
continuous choices. In addition, one may notice the discrepancy (about $3\sim4$ MeV at $k=0$ fm$^{-1}$) between
the s.p. potentials obtained by using the \emph{AV}18 and Bonn B potentials.
 The discrepancy turns out to be regardless of the different choices for the
auxiliary s.p. potential and can be traced back to
the totally different analytical structures of the local
potential (\emph{AV}18) and the non-local potential (Bonn B).
The non-locality in the Bonn B potential increases the attractive
strength of the s.p. potential \cite{ba9,mu1}.

In Fig. \ref{fig2} the saturation curves are given as a function of
density in symmetry nuclear matter with the gap and the continuous
choice adopted. For the two-nucleon \emph{AV}18 potential (left
panel), the results are in good agreement with the previous
two-hole line BHF calculation in Ref. \cite{ba8} with the maximum
deviation less than $1$ MeV for density up to $0.5 $ fm$^{-3}$,
both in the gap and the continuous choices.
Under the continuous choice, the results of Fig. \ref{fig2} have already been
obtained in Refs. \cite{tbf3} and \cite{lizh:2008} for the $AV18$ and Bonn B potentials plus the corresponding TBFs.
Here we repeat those results simply for discussing the influence of the different choices for the auxiliary potential.
One may notice that using
the continuous choice tends to give binding energies about $4-6$ MeV larger
than the gap choice, since the continuous choice
incorporates more non-negligible correlation effects in nuclear medium,
which have been confirmed in Refs. \cite{ba2,ba3,ba4,ba6,ba7} by
solving the three-body Bethe-Fadeev equation with a series of
separable potentials and the full \emph{AV}14, \emph{AV}18 two-body
interactions. The similar results hold for the Bonn B potential
(right panel), except that the obtained EoSs are more
attractive due to the non-locality of the Bonn B potential \cite{ba9,mu1}.
\begin{center}
\includegraphics[width=8cm]{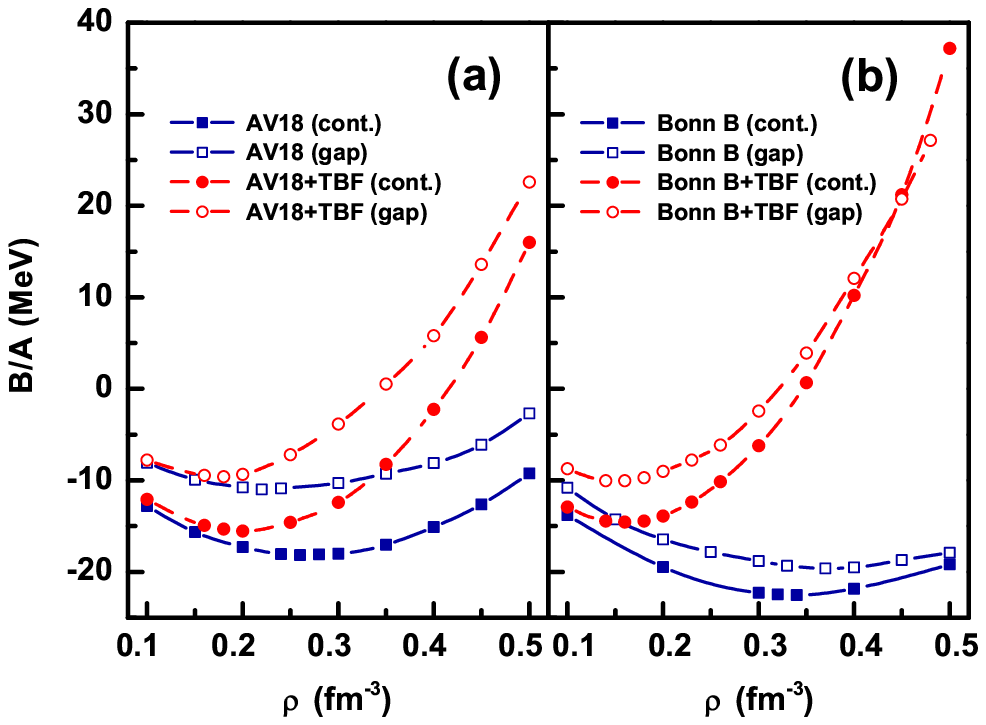}

\figcaption{ (color online) (a) Energy per nucleon
calculated for symmetric nuclear matter with the \emph{AV}18
potential. The lines with circles have been obtained by taking into account the TBF
contribution, while the square symbols denote the results without considering the TBF.  Open and filled
symbols correspond to the two cases with the gap and continuous choice
for the auxiliary potential, respectively. (b) The same as in (a),
but for the Bonn-B potential.}\label{fig2}
\end{center}

On the standard BHF level by adopting purely the \emph{AV}18 and Bonn B
two-body potentials, one obtains too strong binding (solid curves
in Fig. \ref{fig2}). For comparison, the TBF effects are also displayed in the
same figure. The TBFs adopted are self-consistently determined by the \emph{AV}18 and
Bonn B two-body potentials, respectively in Refs. \cite{tbf3,lizh:2008}. The TBF gives a repulsive
contribution to the EoSs (dashed curves) both in the gap and the
continuous choices. The TBF effect is fairly small at low
densities. The repulsive contribution of the TBF increase rapidly as a function of density,
which leads to a stiffer EoS and less binding energies at the
saturation point as compared with the results by using purely the two-body interactions,
under both the gap and continuous choices.
\begin{center}
\includegraphics[width=8cm]{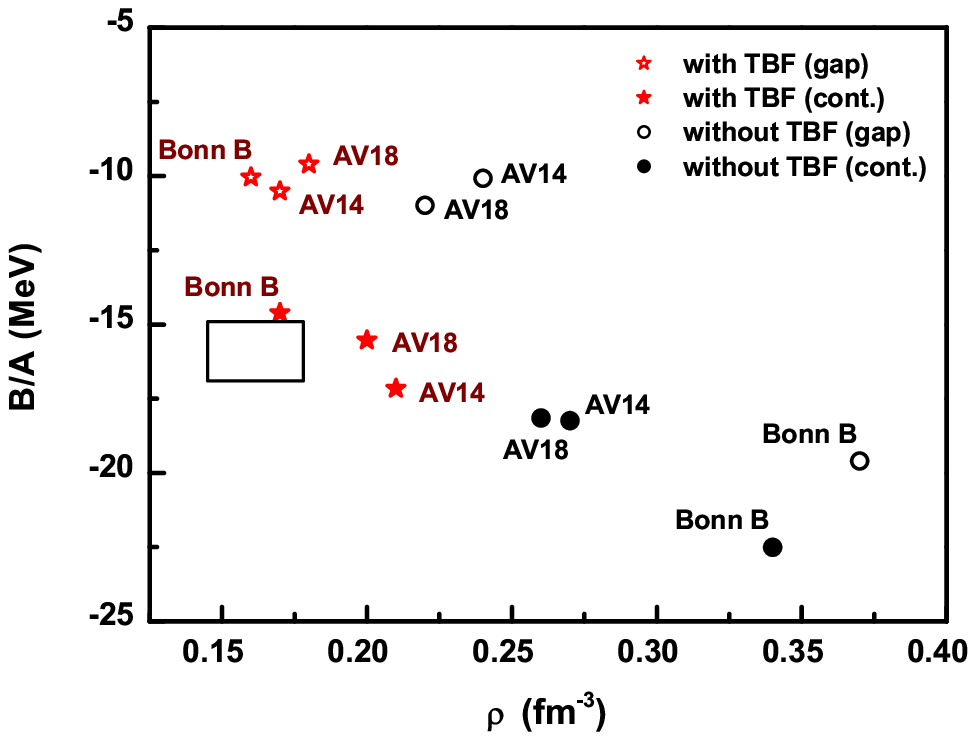}
\figcaption{ (color online) Saturation points obtained
by adopting different NN interactions under the
gap and continuous choices for the auxiliary potential. The square
indicates the empirical saturation region.}
\label{fig3}\end{center}

The saturation points extracted from the previous results are reported in
Fig. \ref{fig3}. Moreover, we give additional results with the
\emph{AV}14 potential in the same figure for the sake of
comparison. The results can be roughly classified into four cases,
i.e., including TBF and excluding TBF
respectively under the the two different kinds of choices (the gap choice and the continuous choice).
By adopting purely the \emph{AV}14, \emph{AV}18, and
Bonn B two-body potentials, the predicted saturation points clearly miss the
empirical saturation region under both the gap and continuous choices. The saturation
densities and the saturation energies obtained within the continuous choice are shown to be much larger than the empirical values.
Similar results are applicable for the gap choice, but the saturation
energies calculated using the \emph{AV}14 and the
\emph{AV}18 potentials are slightly less than the empirical value.
The above results are consistent with the conclusion of the ``Coester
band" both in the gap choice \cite{ba1,co} and the continuous
choice \cite{ba1,li1}. After including the TBFs (with the \emph{AV}14, \emph{AV}18,
and Bonn B potentials), the predicted saturation properties of
symmetry nuclear matter improves remarkably by the TBF
repulsion, especially in the continuous choice. For the gap choice, the TBFs lead to a significant improvement of the saturation density,
whereas the saturation energies turn out to be too small.
It is clearly seen that, after including the TBFs in the
calculation, the predicted saturation points turn to be much
closer to the empirical one in the continuous choice than the
gap choice.
\begin{center}
\tabcaption{\label{tab1} Potential energy per nucleon obtained
for symmetric nuclear matter with the \emph{AV}18 and Bonn B
potentials in the gap and continuous choice, at
$\rho=0.17$fm$^{-3}$. Units are given in MeV.}
\footnotesize
\begin{tabular*}{80mm}{ccccc}
 \toprule &\multicolumn{2}{c}{\emph{AV}18}&\multicolumn{2}{c}{\emph{AV}18 + TBF}\\
 Channel & Gap & Cont. & Gap & Cont. \\
\hline\\
$^{1}S_{0}$ &-15.87& -16.31& -13.05& -13.61\\
$^{3}S_{1}-^{3}D_{1}$ & -15.86& -19.85& -16.82& -20.59\\
$^{3}P_{0}$ & -3.37& -3.44 & -2.09 & -2.11\\
$^{3}P_{1}$ & 10.25& 9.88  & 7.65  &  7.15\\
$^{1}P_{1}$ & 3.95 & 3.87  & 3.34  &  3.28\\
$^{1}D_{2}$ & -2.67& -2.71 & -2.05 & -2.08\\
$^{3}D_{2}$ & -3.87& -3.97 & -1.87 & -1.88\\
$^{3}P_{2}-^{3}F_{2}$ & -7.76 & -8.15 & -7.32& -7.76\\
$J\geqslant3$   & 1.21  & 1.13  & -0.39& -0.47\\\\
Kinetic energy  & 23.04 & 23.04 & 23.04& 23.04\\
Total binding   & -10.94& -16.5 & -9.54& -15.03\\
\hline\\
 &\multicolumn{2}{c}{Bonn B}&\multicolumn{2}{c}{Bonn B + TBF}\\
 Channel & Gap & Cont. & Gap & Cont. \\
\hline\\
$^{1}S_{0}$ &-16.67&-16.88&-14.65&-15.61\\
$^{3}S_{1}-^{3}D_{1}$ & -17.3& -20.52& -18.52& -21.18\\
$^{3}P_{0}$ & -3.57& -3.61& -0.75& -0.34\\
$^{3}P_{1}$ & 10.57& 10.21& 5.23 & 4.25 \\
$^{1}P_{1}$ & 1.52 & 2.36 & 4.01 & 3.45 \\
$^{1}D_{2}$ & -2.43& -2.45& -0.93& -0.79\\
$^{3}D_{2}$ & -4.01& -4.1 & -0.23& 0.07 \\
$^{3}P_{2}-^{3}F_{2}$ & -7.89 & -8.29 & -6.48& -6.45 \\
$J\geqslant3$   & 1.84  & 1.74  & -0.82& -1.04 \\\\
Kinetic energy  & 23.04 & 23.04 & 23.04& 23.04 \\
Total binding   & -14.92& -18.49& -10.1& -14.62\\
\bottomrule
\end{tabular*}
\end{center}

In Table. I, we compare the contributions from various partial wave channels
to the potential energy of symmetry nuclear matter in the gap
choice and the continuous choice at
$\rho=0.17$fm$^{-3}$. The $S$ channel plays a dominate role, while
the contributions from the $P$ and $D$ channels nearly cancel with each
other in the total potential energy when the \emph{AV}18 and Bonn B
two-body potentials plus their corresponding TBFs are adopted. A
similar result has been reported in Ref. \cite{jp1} by adopting the
optical-model with the Reid hard core interaction. By comparing
the results using the two different choices of the auxiliary s. p. potential,
one may notice that the discrepancy between the total binding energies calculated
in the gap and continuous choices mainly comes from the
$^{3}SD_{1}$ channel, in fairly good agreement with the analysis
in Ref. \cite{ba4} by using the \emph{AV}14 potential. It is
seen from Table. I that, after including the TBF, the dominate effect of the
$T=0$ $SD$ coupled channel on the discrepancy remains the same as
that obtained by adopting purely the \emph{AV}18 and Bonn B
two-body potentials.
\begin{center}
\includegraphics[width=8cm]{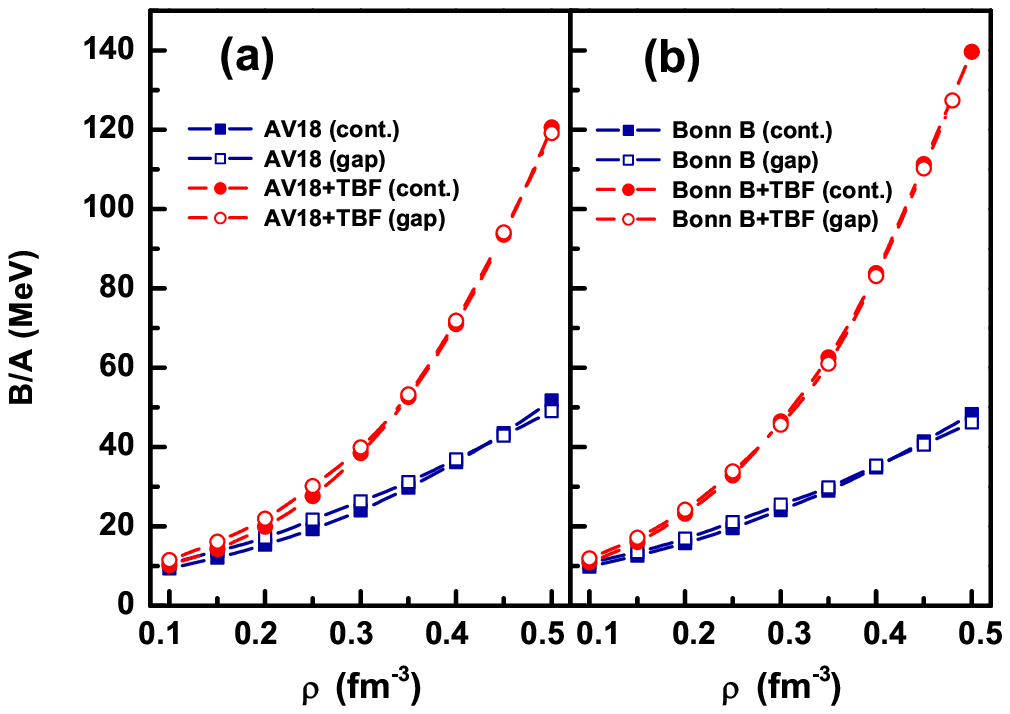}
\figcaption{\label{fig4} (Color online) The same as Fig.2, but the
results are obtained for pure neutron matter.}
\end{center}

In Fig. \ref{fig4}, the EoSs of pure neutron matter obtained within
the gap and the continuous choices is reported. Due to lack of the strongly
attractive effect of the tensor coupling in the isospin $T=0$ $SD$
channel, a faster convergency of the hole-line expansion is expected for pure neutron
matter than that for symmetric nuclear matter. It is seen that, there is a weak dependence on the choice
of the auxiliary s.p. potential by using the \emph{AV}18 and
Bonn B two-body potentials (solid curves). The continuous choice
tends to give a slightly larger binding energies than the gap
choice with the maximum deviation about $2$ MeV, which is much
less than that in the case of symmetry nuclear matter ($4-6$ MeV). This
result is in agreement with Ref.\cite{ba5} and the smaller
average depletion in the neutron Fermi sea\cite{z}. After including
the TBF, the convergent property remains the same
as that obtained using purely the \emph{AV}18 and Bonn B two-body
potentials, whereas the EoSs become much stiffer at high
densities.
\begin{center}
\includegraphics[width=8cm]{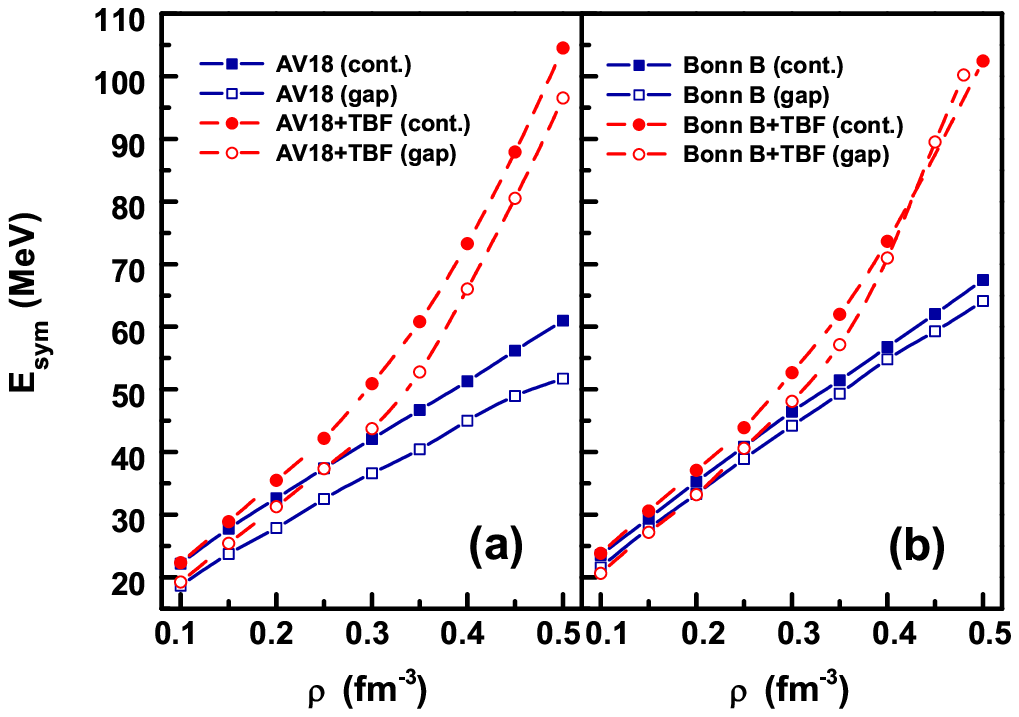}
\figcaption{\label{fig5} (color online) (a) The symmetry energy
obtained by adopting the \emph{AV}18 two-body potential (blue curves with
squares)
and the \emph{AV}18 potential plus the TBF (red curves with circles).
 Open and filled symbols correspond to the two cases with
the gap and continuous choice for the auxiliary potential,
respectively. (b) The same as in (a), but for the Bonn-B
potential.}
\end{center}

According to the microscopic investigations\cite{tbf3,bombaci:1991,zuo:1999},
nuclear symmetry energy can be calculated as the difference between the
 energy per nucleon of pure neutron matter and that of symmetry nuclear matter.
Based on the results in Figs.\ref{fig2} and \ref{fig4}, the symmetry energy is expected to depend on
different choices of the auxiliary s.p. potential. Fig. \ref{fig5} displays the symmetry energies obtained
within the gap and the continuous choices, respectively.
Under the continuous choice, the results in Fig. \ref{fig5} have already been obtained and given in Refs. \cite{tbf3} and \cite{lizh:2008a}.
For all the cases presented in Fig.\ref{fig5}, the symmetry energy increases monotonically
as increasing density, and becomes much stiffer at high
densities when the TBF is self-consistently included in
the calculation. The discrepancy between the calculated EoSs under the gap choice and the
continuous choice, which is sizeable for symmetric matter, is
substantially small for pure neutron matter. Consequently, the
symmetry energy obtained under the gap
choice turns out to be smaller than that under the continuous choice with both the \emph{AV}18 and Bonn B
two-body potentials. The similar results hold after
including the TBFs. For the Bonn B potential, the discrepancy
between of the symmetry energies calculated in the gap choice and the continuous choice, is
quite small, especially at high densities.

\section{Conclusions}

We have studied and compared the TBF effects on the properties of symmetric nuclear matter and pure neutron matter
under the two different choices (i.e., the gap choice and the continuous choice) for the self-consistent auxiliary potential
within the BHF approach.
Special attention has been paid to discuss the difference between the TBF effects under the two different choices.
In our calculation, the recent versions of the realistic \emph{AV}18 and Bonn B two-body potentials plus the corresponding
self-consistent microscopic TBFs have been employed.
By adopting purely the two-body forces, symmetric nuclear matter turns out to be more binding under
the continuous choice than that under the gap choice, confirming the previous results.
Under both the gap and the continuous choices, the TBF effects on the EoSs of symmetric nuclear matter and pure neutron matter are repulsive. The TBF repulsive contribution increases rapidly as a function of density and leads to a significant improvement of the predicted nuclear saturation properties. Therefore, TBF are necessary under both the gap and continuous choices for predicting reliably the EoS of nuclear matter and for reproducing the empirical saturation properties of nuclear matter within the framework of the BHF approach.
Under the continuous choice, the calculated saturation points by including the TBF are closer to the empirical one as compared with the predictions under the gap choice.
After including the TBF,
the convergent property remains the same as that for the two-body interactions. The TBF also affects considerably the auxiliary s.p. potentials in relatively low momentum region, and its effect is to make the s.p. potential less attractive. Furthermore, the symmetry energy obtained under the gap
choice is shown to be slightly smaller than that under the continuous choice. Under both the gap and the continuous choices, the TBF play an important role in determining the high-density behavior of symmetry energy and its effect leads to a strongly stiffening of symmetry energy at high densities.

For the future, the TBF should be included at the three-hole line
level for checking the convergency again by solving the
Bethe-Fadeev equations.
\\[5mm]

\acknowledgments{The work is supported by the National Natural Science
Foundation of China (11175219), the 973 Program of China under No.2013CB834405, the
Knowledge Innovation Project (KJCX2-EW-N01) of Chinese Academy of
Sciences, China}
\\[5mm]

\end{multicols}
\end{CJK*}

\end{document}